\documentclass[a4paper,11pt]{article}
\usepackage{pos}

\title{Scalar and tensor meson dominance and gravitational form factors
  of the pion\footnote{Supported by 
Spanish MINECO and European FEDER funds grant and Project No.
PID2023-147072NB-I00 funded by MCIN/AEI/10.13039/501100011033,
and by the Junta de Andaluc\'\i a grant FQM-225.
}}
 \ShortTitle{Meson dominance of pion gravitational form factors}

\author*[a,b]{Enrique Ruiz Arriola}

\author[c,d]{Wojciech Broniowski}

\affiliation[a]{
Departamento de Fisica Atomica, Molecular y Nuclear , 
Universidad de Granada, E-18071,
Granada, Spain}

\affiliation[b]{
Instituto Carlos I de Fisica Teorica y Computacional, 
Universidad de Granada, E-18071,
Granada, Spain}

\affiliation[c]{
H. Niewodniczanski Institute of Nuclear Physics, PAN, 31-342, Cracow, Poland}

\affiliation[d]{Institute of Physics, Jan Kochanowski University, 
  25-406, Kielce, Poland}

\emailAdd{earriola@ugr.es}
\emailAdd{Wojciech.Broniowski@ifj.edu.pl}

\abstract{We analyze the recent MIT lattice data for the gravitational form factors (GFFs) of the pion which extend up to $Q^2= 2~{\rm GeV}^2$ for $m_\pi=170$~MeV~\cite{Hackett:2023nkr}. We show that simple monopole fits comply with the old idea of meson dominance. We use Chiral Perturbation theory ($\chi$PT) to
  next-to-leading order (NLO) to  transform the MIT data to the physical world with $m_\pi=140~$MeV and
  find that the spin-0 GFF is effectively saturated with the $f_0(600)$
  and the spin-2 with the $f_2(1270)$,  with monopole masses
  $m_\sigma= 630(60)$~MeV and $m_{f_2}= 1270(40)$~MeV. We 
  determine in passing the chiral low energy constants (LECs) from the MIT lattice data alone 
  \[
  10^3 \cdot L_{11} (m_\rho^2)=1.06(15) \, , \qquad 10^3 \cdot L_{12} (m_\rho^2)= -2.2(1) \, ,
  \qquad 10^3 \cdot L_{13} (m_\rho^2) = -0.7(1.1).
  \]
  which agree in sign and order of magnitude 
   with the original estimates by Donoghue and Leutwyler.
   The corresponding D-term (druck) has the value $  
    D(0) = -0.95(3) $. 
   We also analyze the sum
   rules based on perturbative QCD (pQCD) that imply that the
   corresponding spectral functions are not positive definite.  We
   show that these sum rules are strongly violated in a variety of
   $\pi\pi-K \bar K$ coupled channel Omn\`es-Muskhelishvili
   calculations. This is not mended by the inclusion of the pQCD
   tail, suggesting the need for an extra negative spectral strength.  
   Using a simple model implementing
   all sum rules, we find the expected onset of pQCD at very high momenta.  }

\FullConference{10th International Conference on Quarks and Nuclear Physics (QNP2024)\\
8-12 July, 2024\\
Barcelona, Spain\\}

\begin{document}
\maketitle

\section{Introduction}

The stress-energy-momentum tensor (SEM) is the conserved Noether current
corresponding to the symmetry under the space-time translations $x^\mu \to
x^\mu + \epsilon^\mu$. How can we measure it? Hilbert's
proposal was to couple gravity via a curved space time and obtain the
change of the action under an infinitesimal variation (a
microearthquake) around the flat limit, % (see, e.g.,~\cite{Weinberg:1972kfs}),
\begin{eqnarray}
\Theta^{\mu\nu} (x) = \Theta^{\mu \nu} = \frac{-2}{\sqrt{-g}} \frac{\delta S}{\delta g_{\mu \nu}} \Big|_{g^{\mu\nu}= \eta^{\mu\nu}} \, , \qquad \eta^{\mu\nu} = {\rm diag} (1,-1,-1,-1) 
\implies \Theta^{\mu \nu} = \Theta^{\nu \mu} \, . 
\end{eqnarray}
This Hilbert SEM is conserved
and symmetric 
  \begin{eqnarray}
  \Theta^{\mu \nu} = \Theta^{\nu \mu}  \, , \qquad \partial_\mu \Theta^{\mu\nu} =0 \, .
  \end{eqnarray}
Under the Lorentz transformation $ x^\mu \to \Lambda^\mu_\alpha
x^\alpha$ it transforms covariantly, $\Theta^{\mu\nu} (x) \to
\Lambda^\mu_\alpha \Lambda^\nu_\beta \Theta^{\alpha\beta}
(\Lambda^{-1} x)$, but not irreducibly. A naive decomposition into a
traceless and traceful pieces,
\begin{eqnarray}
\Theta^{\mu\nu} = \Theta_S^{\mu \nu} + \Theta_T^{\mu \nu}   \equiv \frac14 g^{\mu\nu} \Theta + \left[ \Theta^{\mu\nu} - \frac14 g^{\mu\nu} \Theta \right] \implies
\partial_\mu \Theta_S^{\mu \nu} = \partial^\nu \Theta \neq 0 \, ,  
\end{eqnarray}
{\it does not} comply to separate conservation.  A consistent decomposition where the  two components are conserved separately was proposed by Raman long ago~\cite{Raman:1971jg}: $
\Theta^{\mu \nu} = \Theta_S^{\mu \nu} + \Theta_T^{\mu \nu}   $, 
with
\begin{eqnarray}
\Theta_S^{\mu \nu} = \frac13 \left[g^{\mu \nu} - \frac{\partial^\mu \partial^\nu }{\partial^2} \right] \Theta \implies  \partial_\mu \Theta_S^{\mu \nu} =0 \, . 
\end{eqnarray}
In QCD, the trace is related to the anomalous divergence of the dilation current 
and the corresponding trace anomaly reads
\begin{eqnarray}
\Theta =\frac{\beta(\alpha)}{4\alpha} {G^{\mu \nu}}^2+ [1+\gamma_m(\alpha)] \sum_f m_f \bar{q}_f q_f, \label{eq:anom}
\end{eqnarray} 
where $\beta(\alpha) = \mu^2 d \alpha / d \mu^2 = - \alpha [\beta_0 (\alpha/4\pi) +
%\beta_1 (\alpha/4\pi)^2 +
  {\cal O}(\alpha^2)  ] < 0 $ is the QCD beta function with $\beta_0 = (11 N_c- 2N_F)/3$, 
  $N_c$ is the number of colors, $N_F$ is the number of flavors, 
$\gamma_m(\alpha)= 2 \alpha/\pi +{\cal O}(\alpha^2) $ is the quark mass anomalous dimension, and  $f$ enumerates the active flavors. One has 
$\alpha(t) = (4\pi /\beta_0) / \ln \left (-t/\Lambda_{\rm QCD}^2 \right)
$ with $\alpha(t) $ real for $t=-Q^2 <0$. We take $\Lambda_Q=0.240$~GeV and $N_F=3$. 

\section{Gravitational form factors}

The GFFs were first introduced by Pagels~\cite{PhysRev.144.1250} (for a review and literature see,
e.g.,~\cite{Polyakov:2018zvc})  as a
way to characterize the mass distribution of hadrons. They describe the
matrix element of SEM
between on-shell states, which for the pion reads
\begin{eqnarray}
&& \langle \pi (p') | \Theta^{\mu \nu}(0) | \pi (p) \rangle \equiv \Theta^{\mu \nu} =  2 P^\mu P^\nu A(q^2)  + \frac12 ( q^\mu q^\nu - g^{\mu \nu} q^2 ) D(q^2 ),
\end{eqnarray}
where $P=\frac12 (p+p')$, $q=p'-p$, and $t=q^2$. On-shell, one has $P^2 = m_\pi^2 - q^2/4$ and $P \cdot q=0$. 
We omit for brevity the isospin indices of the pion, as the considered operator is isoscalar.
The rank-two tensor $\Theta^{\mu \nu}$ can be decomposed into a sum of two separately conserved irreducible tensors corresponding to a
well-defined total angular momentum,  $J^{PC}=0^{++}$ (scalar) and $2^{++}$ (tensor), namely~\cite{Raman:1971jg}
\begin{eqnarray}
  \Theta^{\mu \nu} =   \Theta_S^{\mu \nu}+\Theta_T^{\mu \nu} \, , 
  \qquad \begin{cases}  \Theta_S^{\mu \nu} &= \frac13 Q^{\mu \nu} \Theta \\
    \Theta_T^{\mu \nu} &= \Theta^{\mu \nu}- \frac13  Q^{\mu \nu} \Theta  = 2 \left[ P^\mu P^\nu - \frac{P^2}3  Q^{\mu \nu} \right] A,
  \end{cases}
\end{eqnarray} 
where $Q^{\mu \nu}\equiv g^{\mu \nu}-{q^\mu q^\nu}/{q^2}$.
Since $\Theta$ and $A$ correspond to good $J^{PC}$ channels, they should be regarded as the primary objects, whereas the $D$
form factor mixes the quantum numbers and reads 
\begin{eqnarray}
D= -\frac{2}{3t} \left [ \Theta - \left ( 2 m_\pi^2 -\tfrac{1}{2}\, t \right ) A \right]. \label{eq:Drel}
\end{eqnarray}
The GFFs have eluded direct experimental evaluation except for an extraction for the pion
inferred from the $\gamma \gamma^\ast \to \pi^0 \pi^0$ experimental
data in Ref.~\cite{Kumano:2017lhr}, using the link of GFFs to the generalized
distribution amplitudes (GDA). The first {\it ab initio} calculation by
Brommel~\cite{QCDSF:2007ifr} (with a rather noisy signal)
was analyzed within chiral quark
models (see e.g.~\cite{Broniowski:2008hx}) and in the large
$N_c$ limit~\cite{Masjuan:2012sk}. The present communication is based on our recent work~\cite{Broniowski:2024oyk}, 
where the recent MIT lattice data have been analyzed~\cite{Hackett:2023nkr} in view of the Raman decomposition. 

\section{Analytic properties}

Many properties of $\Theta (s)$ and $A (s)$ are shared with other form factors.
%, and review them here in the single channel case for simplicity. 
The
normalization is %conditions are
  \begin{eqnarray}
\langle \pi(p) | \Theta^{\mu \nu} | \pi(p) \rangle= 2 p^\mu p^\nu \implies   \Theta (0) = 2m_\pi^2 \, , \qquad  A(0)=1 \, . 
  \end{eqnarray}
  $\Theta (s)$ and $A(s)$ are real functions below threshold $s \le 4
  m_\pi^2$ and are analytic in the complex $s$ plane with a
  cut at $s>4m_\pi^2$, so that the discontinuity along the cut is
  purely imaginary, $ {\rm Disc} \Theta(s) \equiv \Theta(s+ i \epsilon)
  - \Theta(s- i \epsilon) = 2 i {\rm Im } \Theta(s+ i \epsilon) $.
  Regarding this branch cut, the form factor has two Riemann sheets
  given by the analytic continuation of their boundary values
  $\Theta^{\rm I}(s) \equiv \Theta(s+i\epsilon)$ and $\Theta^{\rm
    II}(s) \equiv \Theta(s-i\epsilon)$. Further unitarity cuts, at $s= 416m_\pi^2$, 
  $s= 4m_K^2$, \dots , generate new sheets and open subsequent channels.
   
If $T_{IJ}(s)= (e^{2i \delta_{IJ}(s)}-1)/(2i \rho(s))$ denotes the $IJ$ elastic $\pi\pi$
  scattering amplitude for $4 m_\pi^2 \le s \le 16 m_\pi^2$, $\delta_{IJ}(s)$ the phase-shift, and $\rho(s)=\sqrt{1-4m_\pi^2/s}$ the 2-body phase space, then
  Watson's theorem implies 
  \begin{eqnarray}
    {\rm Im} \Theta (s) =  \rho(s) \Theta (s)^* T_{00}(s)  \, , \qquad
        {\rm Im} A (s) =  \rho(s) A (s)^* T_{02}(s)  \, . 
  \end{eqnarray}
With $S^{\rm I} (s)= 1 + 2 i \rho(s) T^{\rm I}(s) $ and $S^{\rm II}(s)=1/S^{\rm I} (s)$ one gets $
\Theta^{\rm II}(s) = S^{\rm II}(s) \Theta^{\rm I}(s)$, hence the resonances correspond to
second Riemann sheet poles of both the form factor and the scattering amplitude.
Moreover, one has the chiral
low-energy theorems~\cite{Novikov:1980fa,Donoghue:1991qv},
  \begin{eqnarray}
    D(0)=-1+ {\cal O}(m_\pi^2/f_\pi^2), \;\;\; \Theta'(0)= 1+ {\cal O}(m_\pi^2/f_\pi^2) \, .  \label{eq:cons}
    \end{eqnarray}
The asymptotic pQCD behavior~\cite{Tong:2021ctu,Tong:2022zax} can be readily
  obtained from Eqs.~(5,6) in~\cite{Tong:2021ctu} by using the
  asymptotic light cone pion wave function $\phi(x) = \sqrt{6} f_\pi x(1-x)$, yielding  
  \begin{eqnarray}
  A(t)=-3 D(t)  \left( 1\! +\! {\cal O} (\alpha) \right) = -\frac{48 \pi \alpha(t) f_\pi^2} {t} \left( 1 \!+\! {\cal O} (\alpha) \right), 
  \label{eq:AD-asymp}
  \end{eqnarray}
and the gluon contribution to the trace anomaly is
  \begin{eqnarray}
  && \hspace{-7mm}\langle \pi (p') | \frac{\beta(\alpha)}{4 \alpha} {G^{\mu \nu}}^2(0) | \pi(p) \rangle  = 16 \pi \beta \left (\alpha(t) \right) f_\pi^2 + {\cal} O(\alpha^3)  = -4 \beta_0 \alpha(t)^2 f_\pi^2 + {\cal} O(\alpha^3).
    \label{eq:Theta-asymp}
  \end{eqnarray}
(confirmed in~\cite{Liu:2024vkj}). 

\section{Dispersion relations, sum rules, and meson dominance}

With all these conditions one has the subtracted dispersion relations
\begin{eqnarray}
  \Theta(t) = 2m_\pi^2+ t \int_{4 m_\pi^2}^{\infty} \!\!\! \frac{ds}{s} \, \frac{{\rm Im}\, \Theta (s)}{s-t}\, , \qquad
  A(t)=1 + t \int_{4 m_\pi^2}^{\infty} \!\!\! \frac{ds}{s} \, \frac{{\rm Im}\, A (s)}{s-t}\,, \label{eq:dr}
\end{eqnarray}
with the sum rules obtained from the vanishing at large $t$
\begin{eqnarray}
  2 m_\pi^2 =  \frac1{\pi} \int_{4 m_\pi^2}^\infty ds \frac{{\rm Im}\, \Theta (s)}{s} \, , \qquad  1=\frac1{\pi} \int_{4 m_\pi^2}^\infty ds \frac{1}{s}{\rm Im} \,A(s),   \, , \qquad  0=\frac1{\pi} \int_{4 m_\pi^2}^\infty ds {\rm Im} \,A(s)   \, ,  
\end{eqnarray}
where clearly ${\rm Im} \,A(s) $ must change sign. With the 
slope at $t=0$ we get
\begin{eqnarray}
2 m_\pi^2 (1\!-\!2 \Theta'(0)) &=& \frac1{\pi} \int_{4 m_\pi^2}^\infty \!\!\! ds  (s-4m_\pi^2)  \frac{{\rm Im} \,\Theta(s)}{s^2} ,
\end{eqnarray}
so ${\rm Im} \Theta(s)$ must also change sign, since $\tfrac{1}{2} < \Theta'(0)= 1
+ {\cal O} (m_\pi^2/f_\pi^2)$. This implies at least one
zero of ${\rm Im} \,\Theta(s)$ and ${\rm Im} \,A(s)$. Indeed, from pQCD
one has, after the analytic continuation $Q^2 \to e^{-i \pi} s$,
\begin{eqnarray}
  \frac1{\pi}  {\rm Im}\, \Theta (s) = 
  - \left(\frac{4 \pi}{\beta_0}\right)^2 \frac{4  \beta_0 L f_\pi^2}{(L^2+\pi^2)^2} + {\cal O} (\alpha^3) \, , \qquad
    \frac1{\pi}  {\rm Im}\, A (s) = 
    - \left(\frac{4 \pi}{\beta_0}\right) \frac{48 \pi  f_\pi^2}{ s(L^2+\pi^2)} + {\cal O} (\alpha^2),
\label{eq:ImpQCD}    
\end{eqnarray}
with $L= \log s/\Lambda_{\rm QCD}^2$, which is manifestly negative. On the other hand, 
in the elastic region
  \begin{eqnarray}
    {\rm Im} \Theta (s) =   |\Theta (s)| \sin \delta_{00} (s) \, , \qquad
        {\rm Im} A (s) = |A (s)| \sin \delta_{02} (s) \, , 
  \end{eqnarray}
  which ensures positivity for attractive interactions for $4 m_\pi^2 < s < 4 m_K^2$, where  
  $0 < \delta_{00}, \delta_{02} < \pi$ 

As mentioned before, the poles of the form factors coincide with the
resonances of the scattering amplitude, which will typically produce a
peak in the spectral function located at $s=M_R^2$ with a width
$\Gamma_R$.  While the shape of the resonance profile in the time-like
region may be complicated by the background and requires abundant
data, its impact in the space-like region is rather mild, such that one
may effectively replace the line shape by a delta function $\delta
(s-M_R^2)$, which in the space-like region turns into an effective
monopole, $ \tilde M_R^2/(\tilde M_R^2+Q^2)$, over a finite $Q^2$
range~\cite{Masjuan:2012sk}.\footnote{This can be seen explicitly considering the form
  factor $F(s)=Z/(M^2-s- i \Gamma M \sqrt{(s-4m^2)/(M^2-4m^2)}$ with $F(0)=1$.
  describing the resonating scattering phase $\delta(s) = {\rm Arg} ( F(s)) $ which
      for $M=0.8$~GeV and $\Gamma=0.7$~GeV is approximated by a monopole with
      an effective mass $\tilde M=0.65$~GeV in the range $Q^2 \le 2~{\rm
        GeV^2}$. }  This provides a superposition of monopole form
    factors with scalar and tensor states,
\begin{eqnarray}
\Theta(t)= 2m_\pi^2 + t \sum_S Z_S \frac{M_S^2}{M_S^2-t} \, , \qquad
A(t)= 1 + t \sum_T Z_T \frac{1}{M_T^2-t} \, , \qquad
\end{eqnarray}
which fulfill the sum rules $2 m_\pi^2 =\sum_S Z_S M_S^2 $, $1 +
{\cal O}(m_\pi^2/f_\pi^2) = \sum_S Z_S $ and $0 =\sum_T Z_T M_T^2 $, as
well as $1 =\sum_T Z_T $ and $0 =\sum_T Z_T M_T^2 $. The minimal
number of resonances fulfilling the sum rules is two, where asymptotically $\Theta(t) = {\cal O}(t^{-1})$ and $A(t) =
{\cal O}(t^{-2})$,
which can be put in agreement with pQCD by explicitly including the perturbative tail.

\section{Analysis of MIT lattice data at $m_\pi=170$~MeV.}

The recent
MIT lattice data~\cite{Hackett:2023nkr} yield
$A(-Q^2)$ and $D(-Q^2)$ in the space like
region $0 \le Q^2 \le 2~{\rm GeV}^2$ at an unphysical {\it fixed} pion mass
$m_\pi^*=170$~MeV. From Eq.~(\ref{eq:Drel}) we have constructed
$\Theta(-Q^2)$ by simply adding errors in $A$ and $D$ in
quadrature.\footnote{This assumes uncorrelated $A$ and $D$, i.e. $\langle A D \rangle = \langle A
  \rangle \langle D \rangle $, and yields $\Delta \Theta^2 = (3Q^2
  \Delta D/2)^2 + (2m_\pi^2+Q^2/2)^2\Delta A^2 $. Alternatively, one
  may naturally assume  that  spin-0 and spin-2 are independent
  degrees of freedom, $A$ and $\Theta$ are uncorrelated,  such that $\langle A \Theta \rangle =
  \langle A \rangle \langle \Theta \rangle $, in which case $\Delta
  \Theta^2 = (3Q^2 \Delta D/2)^2 - (2m_\pi^2+Q^2/2)^2\Delta A^2 $. However, 
  the results of the two scenarios differ little,  since $|D| \sim |A|$ and the relative factor of 9 in
$D$ with respect to $A$ dominates the errors.}
The single resonance model is
\begin{eqnarray}
  A^* (-Q^2)=\frac{m_{f_2}^{*2}}{m_{f_2}^{*2}+Q^2}\, , \qquad
  \Theta^* (-Q^2)=2m_\pi^{*2} -\frac{m_\sigma^{*2} Q^2}{m_\sigma^{*2}+Q^2}. \label{eq:mdlat}
\end{eqnarray}
The fit yields $m_{f_2}^*=1.24(3)~{\rm GeV}$ and
$m_\sigma^*=0.65(3)~{\rm GeV}$, with $\chi^2/{\rm DOF}=0.8$ for 49
data points.\footnote{The quality of
the scalar channel data does not yet allow for a detailed
analysis, which  without further independent input easily incurs in overfitting with the proliferation of parameters. For instance, 
  adding a prefactor $Z_\sigma^*$, one gets $m_\sigma^*=0.790$~GeV and
$Z_\sigma^*=0.812$ with $\chi^2\simeq 16$, but with a large correlation
  between $Z_\sigma^*$ and $m_\sigma^*$. Likewise, adding an $f_0(980)$ term
also suffers from overfitting.}

\begin{figure}[tb]
\begin{center}
\includegraphics[angle=0,width=0.325 \textwidth]{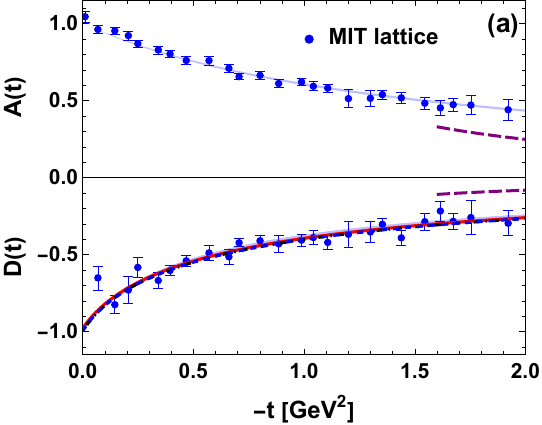} 
\includegraphics[angle=0,width=0.325 \textwidth]{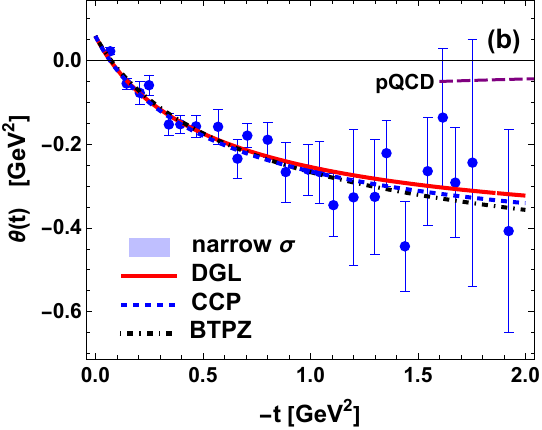}
\includegraphics[angle=0,width=0.325 \textwidth]{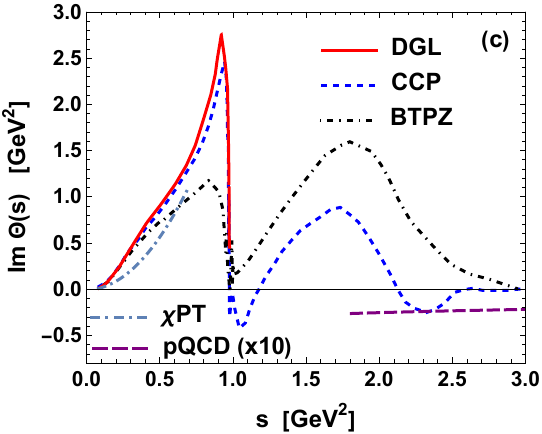} 
\end{center}
\vspace{-5mm}
\caption{GFFs of the pion, plotted as functions of the space-like momentum transfer $-t=Q^2$: (a) Form factors $A(t)$
and $D(t)$ (thin bands, with widths indicating the uncertainty of the fit), compared to the data from~\cite{Hackett:2023nkr} (points with error bars).  (b) The  trace anomaly form factor $\Theta(t)$ (from~\cite{Hackett:2023nkr} via Eq.~(\ref{eq:Drel}), with errors added in quadrature). 
%The tangent (dashed line) at the origin corresponds to the $\chi$PT formula $\theta(t)\simeq 2m_\pi^2+t$. 
The band shows the narrow resonance approximation with $m_{f_2}=1.24(3)$~GeV and
$m_{\sigma}=0.65(3)$~GeV. The model curves are from DGL~\cite{Donoghue:1990xh}, CCP~\cite{Celis:2013xja}, and BPTZ~\cite{Blackstone:2024ouf}. The corresponding  digitized spectral densities used in the dispersion relation~(\ref{eq:dr}) are shown in panel (c).  The long-dashed
lines in (a) and (b) are the asymptotic pQCD results~(\ref{eq:AD-asymp},\ref{eq:Theta-asymp})~\cite{Tong:2022zax} for $\Lambda_{\rm QCD}= 225~{\rm MeV}$; 
the range of the lattice QCD data is far from reaching asymptotics.
The data and the narrow resonance fit correspond to $m_\pi=170$~MeV, whereas DGL, CCP, and BPTZ are at the physical point  $m_\pi=140$~MeV.  \label{fig:fit} }
\end{figure}

\section{Pion mass effects and the chiral low energy constants $L_{11}, L_{12}, L_{13}$}

In QCD, there are only two parameters: $\Lambda_{\rm QCD}$ and the quark
masses $m_q$, which are mapped onto $m_\pi=140$~MeV and $f_\pi=93$~MeV for the simple
two flavor case with equal $m_u=m_d$ assumed all over.
The MIT lattice data are at the unphysical and {\it
  fixed} pion mass value of $m_\pi^*=170$~MeV (and, consequently, at
$f_\pi^*=95$~MeV), so one should account for this effect in
comparison to experimental data.  The only known way to
connect these unphysical data to the real $m_\pi=140$~MeV world is by
means of $\chi$PT, since the corresponding LECs at a
given renormalization scale $\mu$ are independent of the pion
mass. Conventionally, this scale is taken to be about $m_\rho=0.77$~GeV, and transforming the MIT data to the
experimental pion mass $m_\pi=140$~MeV is in general model dependent, except
in the low momentum region where $\chi$PT can be applied. For the absorptive part, one gets perturbatively 
\begin{eqnarray}
  && \hspace{-7mm} \frac1{\pi} {\rm Im}\, \Theta_{\chi} (s) = \sqrt{1-\frac{4 m_\pi^2}{s}} \left(2 m_\pi^2+s\right) \frac{\left(2 s-m_\pi^2\right)}{32 \pi ^2 f_\pi^2}
  + {\cal O} \left(f_\pi^{-4} \right),
\label{eq:ImThetaChPT}
\end{eqnarray}
and hence 
(for {\it given fixed}
$m_\pi$ and $f_\pi$ values) 
\begin{eqnarray}
\Theta_\chi (t) &=& 2 m_\pi^2 + t + \frac{\bar c_1 m_\pi^2 t /2
  - \bar c_2 t^2}{(4 \pi f_\pi)^2} + \frac{t^3}{\pi} \int_{4 m_\pi^2}^\infty \frac{ds}{s^3}\frac{{\rm Im} \Theta_\chi (s)}{s-t} + {\cal O} (f_\pi^{-4}),\label{eq:ThetaChPT} \\ 
  A_\chi (t) &=& 1 - \frac{2 L_{12}}{f_\pi^2} t  + {\cal O} (f_\pi^{-4}).
\label{eq:AChPT}
\end{eqnarray}
From these expressions we get
\begin{eqnarray}
\Theta_\chi'(0)= \frac{\bar c_1 m_\pi^2}{2(4 \pi f_\pi)^2} \, , \qquad
\Theta_\chi''(0)= - \frac{  2 \bar c_2 }{(4 \pi f_\pi)^2} .
\end{eqnarray}
Fitting Eq.~(\ref{eq:ThetaChPT}) to the MIT lattice data for
$Q^2 \le 0.5 {\rm GeV}^2$ yields  $ \Theta_\chi^{*'}(0)=0.87(8) $
and $ \Theta_\chi^{*''} (0)=3.6(4) $, with the correlation $\rho=0.93$ at $m_\pi^{*}=0.170$~GeV
(and $f_\pi^{*}=0.095$~GeV). 

These expressions are good enough to fit to NLO accuracy {\it fixed}
pion mass data. However, in order to relate  {\it different} pion
masses a mass independent renormalization scheme is needed, such as
$\overline{\rm MS}$, as done in~\cite{Donoghue:1991qv}. The
$\chi$PT expression for the SEM with the gravitational LECs,  $L_{11,12,13}$, is  
\begin{eqnarray}
\theta_{\mu \nu}^{(0)} &=& -\eta_{\mu \nu} {\cal L}^{(0)} , \\
\theta_{\mu \nu}^{(2)} &=& \frac{f^2}4 \langle D_\mu U^\dagger D_\nu U
\rangle - \eta_{\mu\nu} {\cal L}^{(2)} ,
\label{eq:en-mom} \\ 
\theta_{\mu \nu}^{(4)} &=& -  \eta_{\mu\nu}{\cal
L}^{(4)} + 2 L_4 \langle D_\mu U^\dagger D_\nu U \rangle \langle
\chi^\dagger U + U^\dagger \chi \rangle \nonumber + L_5 \langle
D_\mu U^\dagger D_\nu U + D_\nu U^\dagger D_\mu U \rangle \langle
\chi^\dagger U + U^\dagger \chi \rangle \nonumber \\ &-& 2
L_{11}\left( \eta_{\mu \nu} \partial^2 - \partial_\mu \partial_\nu
\right) \langle D_\alpha U^\dagger D^\alpha U \rangle -
2 L_{13} \left(\eta_{\mu \nu} \partial^2 - \partial_\mu \partial_\nu
\right) \langle \chi^\dagger U + U^\dagger \chi \rangle \nonumber \\
&-& L_{12} \left( \eta_{\mu\alpha} \eta_{\nu \beta} \partial^2 +
\eta_{\mu\nu} \partial_\alpha \partial_\beta - \eta_{\mu \alpha}
\partial_\nu \partial_\beta - \eta_{\nu \alpha} \partial_\mu
\partial_\beta \right)  \langle D^\alpha
U^\dagger D^\beta U \rangle \,, 
\end{eqnarray} 
where  $\langle . \rangle$ means the trace in the flavor space.
The expansion for the flat space  reads $ 
{\cal L} = {\cal L}^{(0)}+{\cal L}^{(2)} +{\cal L}^{(4)} + \dots $, which contains  the widely known  $L_1,
\dots, L_{10}$ coefficients. 
A direct comparison of Eq.~(\ref{eq:ThetaChPT}) to Ref.~\cite{Donoghue:1991qv} yields 
the identification 
\begin{eqnarray}
    c_1^r=1-128 \pi^2 (6 L_{11}+L_{12}-6 L_{13}) \, , \qquad 
    c_2^r=11/10-64 \pi^2 (3 L_{11}+L_{12}) \, , \label{eq:c1c2}
\end{eqnarray}
where the $K \bar K$ and $\eta \eta$ threshold effects have been neglected.
The coefficients $\bar c_i (m_\pi^2) = c_i^r(\mu^2) + \log
(\mu^2/m_\pi^2) $ are scale independent, whereas $c_i^r(\mu)$ are
independent on the pion mass.  Thus $\bar c_i (4 m_\pi^2) = \bar c_i (
4 m_\pi^{*2}) + \log (m_\pi^{*2}/m_\pi^2)$, and  from the MIT lattice 
data and (\ref{eq:c1c2}) we obtain $  
    D(0) = -0.95(3) $ and  
%Table~\ref{tab:slope}, again with a strong correlation $\rho(\Theta_\chi'(0),\Theta''(0))=0.93 $.  The errors reflect the
% statistical uncertainty only. The final result is then 
\begin{eqnarray}
    10^3 \cdot L_{11} (m_\rho^2)=1.06(15) \, , \qquad 10^3 \cdot L_{12} (m_\rho^2)= -2.2(1) \, ,
  \qquad 10^3 \cdot L_{13} (m_\rho^2) = -0.7(1.1),
\end{eqnarray}
with the strong correlation $\rho(L_{11}, L_{13})=0.93$. Matching the monopole expression of
Eq.~(\ref{eq:mdlat}) to the $\chi$PT results for $m_\pi=140$~MeV yields
$m_\sigma= 630(60)$~MeV and $m_{f_2}= 1270(40)$~MeV.

\begin{figure}[tb]
  \begin{center}
  \includegraphics[angle=0,width=0.325\textwidth]{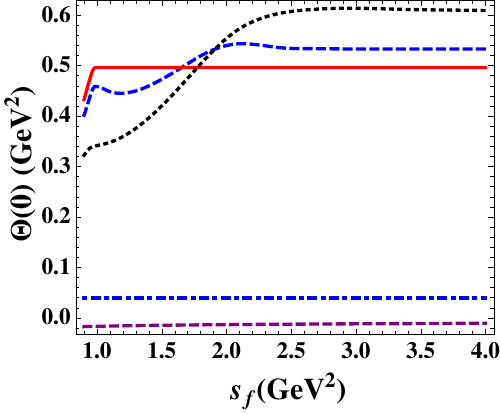}
  \includegraphics[angle=0,width=0.325\textwidth]{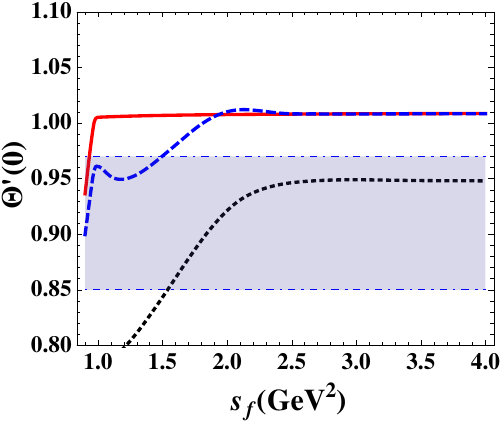}
\includegraphics[angle=0,width=0.325\textwidth]{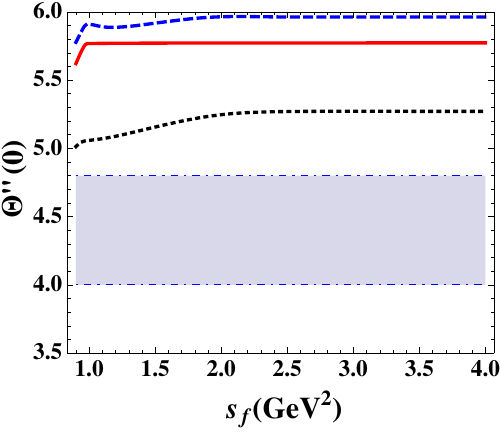} 
\end{center}
\vspace{-5mm}
\caption{Derivatives at the origin,
$  \Theta^{(k)}(0) =  k! \int_{4 m_\pi^2}^\infty ds \, {\rm Im} \, \Theta(s) /s^{k+1}/\pi$, 
  with \mbox{${\rm Im}\, \Theta(s)=  \theta(s_f-s){\rm Im} \, \Theta_f(s) +  \theta(s-s_f){\rm Im}\, \Theta_{\rm pQCD}(s) $},
  plotted as functions of the matching point $s_f$ between
  the DGL~\cite{Donoghue:1990xh}, CCP~\cite{Celis:2013xja} or BPTZ~\cite{Blackstone:2024ouf} $\pi\pi-K \bar K$ coupled channel calculations,
and the  LO pQCD~\cite{Tong:2022zax}, compared with 
our NLO $\chi$PT estimates based on the MIT lattice data extrapolated to the physical  $m_\pi=140$MeV ( $1\sigma$-bands). The tiny and negative dashed line in
panel~(a) is the  pQCD contribution alone (not plotted in (b) or (c)). 
\label{fig:sum-rule}} 
\end{figure}

\begin{table}
\begin{tabular}{|c|c|ccc|ccc|} \hline 
    Ref. & $s_{\rm max}$  & $\Theta (0) $    & $\Theta' (0) $    & $\Theta'' (0) $ & $\Theta (0) $    & $\Theta' (0) $    & $\Theta'' (0) $  \\ 
   &  &  & $ s_f= 1 {\rm GeV}^2 $          &  &  &  $s_f = s_{\rm max} $          &      \\ \hline 
DGL~\cite{Donoghue:1990xh} & $1 {\rm GeV}^2$  & 0.50 & 1.01 & 5.77 & 0.50 & 1.01 & 5.77 \\
CCP \cite{Celis:2013xja}& $3 {\rm GeV}^2$ & 0.46 & 0.97 & 5.91 & 0.53 & 1.01 & 5.96 \\
BPTZ~\cite{Blackstone:2024ouf}  & $4 {\rm GeV}^2$ & 0.34 & 0.79 & 5.06 & 0.61 & 0.95 & 5.27 \\ \hline
$\chi{\rm PT}_{\rm NLO}$~\cite{Broniowski:2024oyk} & $ \infty $ & 0.0392 & 0.91(6) & 4.4(4) & & & \\
$ {\rm Monopole}$~\cite{Broniowski:2024oyk} & $ - $ &  0.40(7) &  1 & 5.2(9) & & & \\ \hline
\end{tabular}
\caption{Sum rule derivatives at the origin, 
  $\Theta_f^{(k)}(0) =  k! \int_{4 m_\pi^2}^{s_f} ds\, {\rm Im} \, \Theta(s) /s^{k+1}/\pi$, 
  for several coupled channel calculations, compared with estimates
  from the MIT lattice data.}
\label{tab:slope}
\end{table}

\section{Sum rules revisited}

Phenomenological studies of the $\pi\pi-K \bar K$
coupled-channel unitary approach based on the Omn\`es-Muskhelishvili
equations indeed confirm a zero of ${\rm Im}\, \Theta (s)$ at about
$s_f \sim 1 {\rm
  GeV}^2$~\cite{Donoghue:1990xh,Celis:2013xja,Blackstone:2024ouf} and
a certain oscillating pattern beyond that scale, see
Fig.~\ref{fig:fit} (c).  Despite their differences in the time-like
region (Fig.~\ref{fig:fit} (c)), they are hardly distinguishable in the
space like region (Fig.~\ref{fig:fit} (b)), where they strongly resemble the
monopole fit with $m_\sigma= 630(60)$~MeV, confirming the mentioned space-like
insensitivity to detailed time-like physics. From the exact dispersion
relations 
\begin{eqnarray}
\Theta^{(k)}(0) = \frac{k!}{\pi} \left( \int_{4 m_\pi^2}^{s_f} + \int_{s_f}^\infty \right) ds \frac{{\rm Im} \, \Theta(s)}{s^{k+1}} \, ,
\end{eqnarray}
for $ k=0,1,2, \dots $,
which are sum rules for derivatives at the origin, increasingly
independent on high energies. The contributions up to the upper limit
$ s_f$ are listed in Table~\ref{tab:slope}. In
Fig.~\ref{fig:sum-rule} we plot the values of $\Theta(0)$, $\Theta'(0)$, and
$\Theta''(0)$  as functions of the matching point $s_f$
between the $\pi\pi-K\bar K$ coupled channel solution and the LO-pQCD. As
we see, the mass sum rule $\Theta(0)=2 m_\pi^2 $ is badly violated,
whereas the $\Theta'(0)$ and $\Theta''(0)$ sum rules overhit the
$\chi$PT estimate, suggesting that
there is a significant missing negative contribution in the higher
energy tail before pQCD sets in.\footnote{This is similar to the pion
  charge form factor $F_Q(t)$ determined from the experimental data for $s_f
  \sim 9~{\rm GeV}^2 $~\cite{RuizArriola:2024gwb}, which fulfills similar
  sum rules as $A(t)$; the large violations happen with pQCD to NNLO
  ~\cite{Sanchez-Puertas:2024siv}.}
 
Among the many possibilities, we assume here a simple  power
model in the intermediate $s$ range which resembles the asymptotic behavior
of a radial Regge spectrum~\cite{RuizArriola:2008sq}
\begin{eqnarray}
  {\rm Im} \Theta_\epsilon (s) =
%  m_\sigma^4 \delta(s-m_\sigma^2) + \theta(\Lambda_p^2 - s)
%  \theta( s-\Lambda^2)
  {\rm Im} \Theta (s_f) \left(\frac{s_f}{s} \right)^\epsilon \, , \qquad
 s_f \le s \le \Lambda_{\rm pQCD}^2 \, . \label{eq:epsi}
%  +\theta(s-\Lambda_p^2) {\rm Im} \Theta_{\rm pQCD}(s)
\end{eqnarray}
The discussed sum rule violations correspond to taking $s_f=\Lambda_{\rm pQCD}^2$. 
Instead, in the limit $\Lambda_{\rm pQCD}^2 \gg s_f$ we may neglect the pQCD tail 
and the  sum rules for a sharp $\sigma$ resonance contribution plus Eq.~(\ref{eq:epsi}) read
  \begin{eqnarray}
    2 m^2 = m_\sigma^2 +  \frac{{\rm Im} \Theta (s_f)}{\pi \epsilon} \, , \quad 
    \Theta'(0) &=& 1 + \frac{{\rm Im} \Theta (s_f)}{\pi (1+\epsilon) s_f} \, , \quad  \Theta''(0) = \frac{2}{m_\sigma^2} + \frac{{\rm Im} \Theta (s_f)}{\pi (2+\epsilon) s_f^2} \, . 
  \end{eqnarray}
  Then $\epsilon >0 \implies {\rm Im}\, \Theta( s_f) < 0 \implies
  \Theta'(0) < 1 $, in agreement with the phenomenological findings
  (see Fig.~\ref{fig:fit} (c)). For simplicity, the scale $s_f = 1~{\rm
    GeV}^2$ is chosen, since $\Theta_f'(0)=1$ (in the DGL scenario).
  Taking $m_\sigma=0.63$~GeV and $\Theta'(0)=0.91$ we get $ {\rm Im}
  \Theta(s) = -0.378/s^{0.33}~{\rm GeV}^2$ and $\Theta''(0)=4.7$, in
  agreement with $\chi$PT (see Table~\ref{tab:slope}). This power
  falls off faster than the pQCD contribution, Eq.~(\ref{eq:ImpQCD}),
  and will eventually intersect when ${\rm Im} \,\Theta_{\rm p QCD}
  (s_p)= {\rm Im} \,\Theta_\epsilon (s_p) $  at the large scale
  $\sqrt{s_p}= 0.6 \cdot 10^6 {\rm GeV}$ (see also
  Re.(\cite{RuizArriola:2008sq}), such that in practice in this model the pQCD
  expression does not contribute significantly to the sum rules. We
  refrain here from a thorough analysis, as the model is only schematic. However,
  at face value it shows that despite of reinforcing the
  usefulness of the meson dominance idea in the intermediate energy
  region, the inclusion of rigorous QCD sum rules for the form factors
  calls for a negative spectral strength much larger than a direct
  extrapolation of pQCD to any energy above $1$~GeV. Simple models
  compatible with low energy phenomenology and fulfilling the sum
  rules suggests a delayed onset of pQCD, moved to very high energies.

\acknowledgments We thank Pablo S\'anchez Puertas for discussions and
Dimitra Pefkou for correspondence.

\providecommand{\href}[2]{#2}\begingroup\raggedright\endgroup

\end{document}